\documentclass[aps,pra,preprint,groupedaddress,floatfix]{revtex4-2}

\usepackage{graphicx}
\usepackage{overpic}

\usepackage[absolute,overlay]{textpos}
\usepackage{array}
\usepackage{booktabs}
\usepackage{soul}
\usepackage{xcolor}
\usepackage[normalem]{ulem}
\usepackage[sfdefault=cmbr]{isomath}
\usepackage{siunitx}
\usepackage[version=4]{mhchem}
\usepackage{braket}

\usepackage[final, breaklinks = true]{hyperref}
\usepackage{cleveref}

\providecommand*{\ped}[1]{\ensuremath{_\textup{#1}}}		%same with text instead of mathrm

\begin{document}

\title{Atomic thermometry in optical lattice clocks: model discrepancies and their impact on lattice light shifts}

\author{Irene Goti}
\email[]{i.goti@inrim.it}
\affiliation{Istituto Nazionale di Ricerca Metrologica (INRIM), 10135 Torino, Italy}

\author{Tommaso Petrucciani}
\affiliation{Istituto Nazionale di Ricerca Metrologica (INRIM), 10135 Torino, Italy}

\author{Stefano Condio}
\affiliation{Istituto Nazionale di Ricerca Metrologica (INRIM), 10135 Torino, Italy}

\author{Filippo Levi}
\affiliation{Istituto Nazionale di Ricerca Metrologica (INRIM), 10135 Torino, Italy}

\author{Davide Calonico}
\affiliation{Istituto Nazionale di Ricerca Metrologica (INRIM), 10135 Torino, Italy}

\author{Marco Pizzocaro}
\email[]{m.pizzocaro@inrim.it}
\affiliation{Istituto Nazionale di Ricerca Metrologica (INRIM), 10135 Torino, Italy}

%\date{\today}

\begin{abstract}
Accurate measurement of atomic temperature is fundamental for a wide range of applications, from quantum sensing to precision metrology. In optical lattice clocks, precise characterization of atomic temperature is required to minimize systematic uncertainties at the $10^{-18}$ level. In this work, we investigate atomic temperature measurements in the ytterbium optical clock developed at INRIM, IT-Yb1, employing sideband and transverse Doppler spectroscopy across a wide range of trapping conditions in a one-dimensional optical lattice. By implementing clock-line-mediated Sisyphus cooling [Phys. Rev. Lett. 133, 053401 (2024)], we reduce the atomic temperature and enable operation at shallower lattice depths down to $D = 50\,E\ped{R}$, where $E\ped{R}$ is the photon recoil energy. We compare temperature estimates obtained from the perturbed harmonic oscillator model with those derived using a Born-Oppenheimer model [Phys. Rev. A 101, 053416 (2020)]. Discrepancies of up to a factor of four are observed between the temperature estimates obtained with the two approximations. We assess the impact of these modeling differences on the evaluation of lattice light shifts and find relative frequency deviations of up to $8\times10^{-17}$. Although sufficiently low atomic temperatures and the consistent use of theoretical models reduce these discrepancies to the low $10^{-18}$ level, these results motivate further investigations that will be relevant to state-of-the-art optical lattice clocks.

%The largest discrepancies occur when different theoretical models are used to determine the temperature and evaluate the lattice light shift. Sufficiently low atomic temperatures reduce these model-dependent discrepancies to the $1\times10^{-18}$ level, which may still limit the accuracy of optical lattice clocks.

\end{abstract}


\maketitle

\section{Introduction}

Optical dipole traps~\cite{Grimm2000} and optical lattices~\cite{Takamoto2005,bloch2005} provide powerful platforms for the control and manipulation of ultracold atoms, enabling applications ranging from quantum sensing~\cite{Charriere2012,panda2024} and quantum simulation~\cite{Gross2017} to high-accuracy atomic clocks~\cite{Ludlow2015} and tests of fundamental physics~\cite{Safronova2018,Takamoto2020}. In many of these applications, accurate measurement of the atomic temperature is an important diagnostic tool~\cite{Hilton2020,Zhou2011,Glatthard2022,Mitchison2020} because it describes the distribution of the atoms over motional states in the trap. In particular, atomic temperature plays a central role in optical lattice clocks, which can nowadays achieve total uncertainties at the level of \num{e-18}~\cite{Beloy2021,Ushijima2015,McGrew2018}. 

In optical lattice clocks, the motional state of the atoms influences the residual lattice light shift. One strategy to mitigate this effect is cooling atoms close to the fundamental motional state, by implementing post-cooling techniques such as sideband cooling, clock-line-mediated Sisyphus cooling, or applying cleaning dips~\cite{Ushijima2018, Nemitz2019, Nemitz2016, Ushijima2015, Falke2014, Chen2024, Zhang2022}. A second strategy, first proposed by Brown et al.~\cite{Brown2017}, exploits the fact that the atomic temperature in the lattice depends on the lattice depth and when this dependence is linear, a considerable simplification of the lattice shift equation can be obtained~\cite{Brown2017, Beloy2020}. 
Moreover, the motional state of the atoms influences the collisional shift in the clock \cite{Lemke2011,Nemitz2019,Yanagimoto2018}.
For these reasons, it is essential to develop an accurate method to measure the temperature of atoms trapped in an optical lattice. 
Several experimental techniques have been developed to measure cold atom temperatures, including time-of-flight imaging~\cite{Lett1988,Salomon1990}, release and recapture~\cite{mudrich2002}, and spectroscopic techniques based on the light shift induced by the trap~\cite{Hilton2020}, such as carrier thermometry~\cite{McDonald2015, Han2018}. The most common technique in 1D optical lattice clocks is sideband spectroscopy~\cite{Blatt2009}: clock spectroscopy is performed along the axis of strong confinement and motional sidebands are well resolved. Fitting the resulting spectrum provides information on the radial and longitudinal temperatures, as well as the trap depth. Sideband spectroscopy has the advantage that it can be carried out in a magic frequency trap without requiring spatially resolved imaging or the need to turn off the trap, enabling in situ thermometry of the trapped atoms. The radial temperature can alternatively be determined using transverse Doppler spectroscopy, in which atoms are probed along the weakly confined axis and the excitation spectrum exhibits a Doppler-broadened profile similar to free-space atoms~\cite{Wineland1979}.

Traditional approaches to analysing sideband spectra~\cite{Blatt2009} rely on the perturbed harmonic oscillator approximation, where atoms are assumed to occupy only a small region near the bottom of the trap. This allows the lattice potential to be expanded in a power series, with higher-order terms treated as small perturbations to the harmonic energy levels. Recently, Beloy et al. introduced the Born-Oppenheimer approximation to describe the atomic motional state and the light shift of atoms trapped in an optical lattice to go beyond the perturbed harmonic oscillator approximation~\cite{Beloy2020, Bothwell2025a}. In molecular theory, the Born-Oppenheimer approximation separates the electronic motion (fast) from the nuclear motion (slow). The electronic wave function is solved for fixed nucleus positions and then the nuclei are considered to evolve on the resulting potential surfaces. Beloy et al. draw a parallel to the atomic motion in a 1D optical lattice where the longitudinal motion (fast) is separated from the radial motion (slow), due to the much higher trapping frequency along the lattice axis compared to the radial directions.

In this paper, we investigate atomic temperature measurements performed with the Yb optical lattice clock IT-Yb1~\cite{Pizzocaro2020, Goti2023} via sideband and transverse Doppler spectroscopy. We consider a wide range of trapping conditions by using the clock-line-mediated Sisyphus cooling technique recently proposed and demonstrated by Chen et al.~\cite{Chen2024}, and we compare results from the perturbed harmonic oscillator and Born-Oppenheimer models. The corresponding impact on the evaluation of the lattice light shift is also calculated. The paper is structured as follows: section II reviews the perturbed harmonic oscillator, Born-Oppenheimer, and Doppler approximations used throughout this work, highlighting their limitations; section III describes the experimental setup of IT-Yb1; section IV presents the results obtained; section V discusses the results and their impact on lattice shifts calculations. Section VI provides the conclusions of our work.

\section{Models}

\subsection{Sideband shape in the perturbed harmonic oscillator approximation}
Here we review the traditional description of the sideband spectrum for atoms trapped in a 1D optical lattice of depth $D$, based on the perturbed harmonic oscillator approximation as presented by Blatt et al.~\cite{Blatt2009} and other studies~\cite{Ludlow2008a,LeTargat2007,Westergaard2010,Nemitz2019}.

The full lattice potential is
\begin{equation}
    U(r,z) = -D e^{-2r^2/w^2}\cos^2(kz),
\end{equation}
where $r=\sqrt{x^2+y^2}$ and $z$ are the radial and longitudinal positions, $w$ is the lattice beam waist
($1/e^2$ intensity radius), $k=2\pi/\lambda$ is the lattice wavevector, and $\lambda$ is the lattice wavelength. In the perturbed harmonic oscillator approximation, the potential is expanded in a series up to the longitudinal quartic distortion and the coupling between longitudinal and transverse motion. Then, in first-order perturbation theory, the atomic motional state in the lattice is described by the quantum numbers $n_x, n_y, n_z$, and the spectrum is:
\begin{multline}\label{eq:1}
    E_{n_x n_y n_z} \simeq -D + h \nu_z \left(n_z + 1/2\right) + h \nu_r (n_x + n_y + 1) - \frac{h \nu\ped{rec}}{2}\left(n_z^2 + n_z + 1/2\right)\\ - h \nu\ped{rec}\frac{\nu_r}{\nu_z}(n_x + n_y + 1) \left(n_z + 1/2\right),
\end{multline}
where $h$ is the Planck constant, $\nu_z = 2\sqrt{D E\ped{R}}/h$ is the longitudinal trap frequency and $\nu_r = \sqrt{D / (m \pi^2 w^2)}$ is the radial trap frequency, $\nu\ped{rec}=h/(2m\lambda^2)$ is the lattice recoil frequency, $m$ is the atomic mass, and $E\ped{R} =h \nu\ped{rec}$ is the lattice photon recoil energy. 
%Equation~\ref{eq:1} retains only the first-order energy corrections arising from the longitudinal quartic distortion and the coupling between longitudinal and transverse motion.

When atoms are driven by a laser with detuning $\delta$ from a resonance $\ket{g} \to \ket{e}$, where $g$ and $e$ are ground and excited states, sidebands appear due to transitions that change the longitudinal vibrational quantum number. The blue sideband is given by transitions where $n_z \rightarrow n_z'=n_z + 1$, while for the red sideband $n_z \rightarrow n_z'=n_z - 1$. 
Assuming the radial quantum numbers $n_x$, $n_y$ are conserved (harmonic-oscillator selection rules), transitions $n_z \rightarrow n_z+1$ happen at the frequency:
\begin{multline}\label{eq:ho-freq}
\delta\nu(n_x, n_y, n_z) = \frac{E_{n_x n_y n_z+1} - E_{n_x n_y n_z}}{h} = \nu_z - \nu\ped{rec}(n_z+1) - \nu\ped{rec}\frac{\nu_r}{\nu_z}(n_x + n_y +1).
\end{multline}

The blue sideband shape in the perturbed harmonic oscillator approximation can be written as a sum of power-broadened, saturated, Lorentzian spectra centered in $\delta\nu(n_x, n_y, n_z)$ for each $n_z \rightarrow n_z+1$ transition:
\begin{equation}\label{eq:blatt-blue}
   \sigma\ped{blue}(\delta) \propto \sum_{n_x n_y n_z} \frac{p_{n_x n_y n_z}}{1+(\delta - \delta\nu)^2/\gamma^2},
\end{equation}
where $p_{n_x n_y n_z}$ is the probability for atoms to be in the $n_x, n_y, n_z$ state and $\gamma$ is the power-broadened linewidth. 

By introducing $n_r = n_x + n_y$ and a degeneracy factor $(n_r+1)$, the probability for atoms to be in the $n_z, n_r$ state can be given as a function of the longitudinal and radial temperatures $T_z$ and $T_r$~\cite{Blatt2009}:
\begin{equation}\label{eq:blatt-temp}
    p_{n_r n_z} \propto (n_r + 1)\exp\left[-h\nu_r(n_r+1)/(k\ped{B}T_r)\right]\exp\left[-E_{0 0 n_z}/(k\ped{B} T_z)\right],
\end{equation}
where $k\ped{B}$ is the Boltzmann constant.
Eq.~\ref{eq:blatt-blue} can be simplified by replacing the sum with the dominant term for each $n_r$, where $\delta = \delta\nu(n_z, n_r)$, resulting in the typical exponential-shape sideband on the side facing the carrier. Similar calculations can be carried out for the red sideband.

\subsection{Sideband shape in the Born-Oppenheimer approximation}
%In the Born-Oppenheimer approximation, atoms are described by the longitudinal quantum number $n_z$, the radial quantum number $n_r$ and the angular momentum quantum number $l$ with the motional wavefunction in cylindrical coordinates factorized as \(\Psi_{n_r l n_z}(r,\phi,z)\simeq (2\pi r)^{-1/2}R_{n_r l n_z}(r)Z_{n_z}(r,z)e^{il\phi}\). \(Z_{n_z}(r,z)\) is obtained by solving the longitudinal motion at fixed $r$, and its eigenvalue \(U_{n_z}(r)\) are the potential curves over which \(R_{n_r l n_z}(r)\) are calculated. The radial potential curves $U_{n_z}(r)$ can be calculated for the perturbed harmonic oscillator potential or for the full lattice potential, which support analytical calculations based on Mathieu functions~\cite{Beloy2020}.

In the Born-Oppenheimer approximation, atoms are described by the longitudinal quantum number $n_z$, the radial quantum number $n_r$ and the angular momentum quantum number $l$. In cylindrical coordinates
($r,\phi,z$), where $r$ and $z$ are the radial and longitudinal positions and $\phi$ is the azimuthal angle, the motional wavefunction is factorized as
\begin{equation}
\Psi_{n_r l n_z}(r,\phi,z)\simeq (2\pi r)^{-1/2}\mathcal{R}_{n_r l n_z}(r)\mathcal{Z}_{n_z}(r,z)e^{il\phi},
\end{equation}
where $\mathcal{Z}_{n_z}(r,z)$ is the longitudinal wavefunction obtained by solving the motion along $z$ at fixed $r$. The corresponding eigenvalue $U_{n_z}(r)$ is a potential curve, and $\mathcal{R}_{n_r l n_z}(r)$ is the radial wavefunction obtained by solving the motion in this potential. The curves $U_{n_z}(r)$ can be calculated using either the perturbed harmonic oscillator potential or the full lattice potential. In the latter case, the longitudinal solutions can be expressed in terms of Mathieu functions~\cite{Beloy2020}.

To describe transitions within the Born–Oppenheimer model, we apply the Franck–Condon principle and the turning-point approximation.
In molecular dynamics, the Franck-Condon principle states that, owing to the much larger masses of the nuclei compared with the electrons, electronic transitions classically conserve the nuclear positions and momenta~\cite{Condon1926,Condon1947, Mulliken1971}.
Depending on the shapes of the initial and final potential-energy surfaces, this generally leads to a band of allowed transition energies. Transition probabilities are proportional to the squared overlap integral between the nuclear wavefunctions in the initial and final electronic states, or Franck-Condon factors~\citep{Szudy1975, Tellinghuisen1987,Beloy2026}. Franck-Condon factors can give rise to complex vibrational structure in molecular~\cite{Mulliken1971,Tellinghuisen1987} and solid-state spectra~\cite{Lax1952}, and are also relevant to pressure-broadened line shapes~\cite{Szudy1975} and cold-molecule photoassociation~\cite{Boisseau2000,Zelevinsky2006}. %For instance, Franck-Condon factors can give rise to complex vibrational structure in molecular~\cite{Mulliken1971, Tellinghuisen1987} and solid-state spectroscopy~\cite{Lax1952}, as well as in pressure-broadened line shapes~\cite{Szudy1975}, and the photoassociation of cold molecules~\cite{Boisseau2000, Zelevinsky2006}.

%A further approximation is required to assign a single representative atomic energy class to a given laser frequency.
%For a laser at a fixed frequency, the most probable transition is assumed between states having the same classical turning points~\cite{Condon1926,Condon1947, Mulliken1971}.
%This turning-point approximation provides a model of manageable complexity by selecting the atomic energy class for which a turning-point transition is expected to provide the dominant contribution.
The turning-point approximation is required to achieve a model of manageable complexity by associating each laser frequency with a single representative atomic energy class, rather than explicitly treating the full set of allowed transitions. In this approximation, the class is selected by assuming that the most probable transition occurs between states with the same classical turning points~\cite{Condon1926,Condon1947,Mulliken1971}. 
%This approximation simplifies the model by considering only the dominant turning-point contribution rather than explicitly treating the full set of allowed transitions.
Classically, this is motivated by the fact that the slow degree of freedom spends more time near the turning points.
In a quantum-mechanical treatment, this can be verified by calculating Franck-Condon factors~\citep{Szudy1975, Tellinghuisen1987}. 
In the semiclassical or Wentzel-Kramers-Brillouin (WKB) approximation, the Franck-Condon factor integrals involve the product of oscillatory terms that accumulate mainly near points of stationary phase, or Condon points, which are the points where the classical nuclear position and momentum are conserved.
Condon points near classical turning points result in higher overlaps and transition probabilities because the nuclear momentum is small, leading to results consistent with the classical picture~\cite{Mulliken1971,Szudy1975,Tellinghuisen1987}.

When applying the Franck–Condon principle and the turning-point approximation to the optical-lattice potential, we assume that transitions conserve the slow degree of freedom, which, in this case, is the radial coordinate $r$.
%For atoms interacting with a laser detuned from resonance by $\delta$,
We assume that the transition at frequency $\delta\nu$ happens at the classical turning point $r\ped{c}$ determined by inverting
\begin{equation}\label{eq:bo-freq}
\delta\nu = [U_{n_z'}(r\ped{c}) - U_{n_z}(r\ped{c})]/h.
\end{equation}
We can neglect the role of the angular momentum~\cite{Tellinghuisen1987} because $l$ is conserved for a Franck-Condon transition and the centrifugal potential~\cite{Beloy2020} will cancel out in Eq.~\ref{eq:bo-freq}.
Moreover, there is a single classical turning point, so this relation fixes a one-to-one relationship between the transition frequency $\delta\nu$ and the classical turning point $r\ped{c}$.
The limitations of this approximation and comparison with Eq.~\ref{eq:ho-freq} are discussed in sections~\ref{sec:comparison} and~\ref{sec:limitations}.

As in the perturbed harmonic oscillator approximation, we can write the blue sideband shape for a probe at detuning $\delta$ as a sum of power-broadened, saturated, Lorentzian profiles, one for each $n_z \rightarrow n'_z = n_z+1$ transition
\begin{equation}\label{eq:beloy-blue}
\sigma\ped{blue}(\delta) \propto \sum_{n_r l n_z} \frac{p_{n_r l n_z}}{1+(\delta - \delta\nu)^2/\gamma^2}.
\end{equation}

%For the population of the motional state $p_{n_r l n_z}$ we assume the generalized two-temperature Boltzmann distribution~\citep{Beloy2020}, which is similar to Eq.~\ref{eq:blatt-temp}:
%\begin{equation}
%p_{n_r l n_z} \propto \exp\left[-(E_{n_r l n_z} -E_{00n_z})/(k\ped{B}T_r)\right] \exp\left[-E_{00n_z}/(k\ped{B}T_z)\right].
%\end{equation}

Following the WKB approximation~\cite{Beloy2020}, the summation over $n_r$ and $l$ in Eq.~\ref{eq:beloy-blue} can be replaced by integration over the energy $E$ and $l$ because the spectrum is dense. Carrying out the integration over $l$ first and introducing the density of states $G_{n_z}(E)$ and the fractional population $P_{n_z}(E)$, we can write the blue sideband shape as:
\begin{equation}\label{eq:vband}
	\sigma\ped{blue}(\delta) \propto \sum_{n_z}\int_E dE \frac{G_{n_z}P_{n_z}}{1+(\delta - \delta\nu)^2/\gamma^2},
\end{equation}
where $G_{n_z}$, $P_{n_z}$, $\delta\nu$ and $\gamma$ can all be calculated from the energy $E$ or the corresponding classical turning point $r\ped{c} = R_{n_z}(E)$, where $R_{n_z}(E)$ is the inverse function of the potential curve $U_{n_z}(r)$. The density of states is $G_{n_z}(E) = \frac{m}{2\hbar ^2}[R_{n_z}(E)]^2 = \frac{m r\ped{c}^2}{2\hbar ^2}$ \cite{Beloy2020}.
For $P_{n_z}(E)$ we assume a generalized two-temperature Boltzmann distribution~\citep{Beloy2020}, similar to Eq.~\ref{eq:blatt-temp}, written as:
\begin{equation}\label{eq:boltzmann}P_{n_z}(E) \propto e^{-(E - U_{n_z}(0))/k\ped{B}T_r} e^{-U_{n_z}(0)/k\ped{B}T_z}.
\end{equation} 
The frequency $\delta\nu$ can be calculated as a function of $r\ped{c}$ with Eq. \ref{eq:bo-freq}.

The power-broadened linewidth $\gamma$ for each transition is proportional to the Rabi frequency of the transition $\Omega$. For trapped atoms, the Rabi frequency associated with a clock laser of wavevector k\ped{c}, propagating along the lattice direction $z$, is:
\begin{equation}
\Omega^2 = |\braket{n_r' l' n_z'|\Omega_0 \exp{i k\ped{c} z}|n_r l n_z}|^2,
\end{equation}
where $\Omega_0$ is the free space Rabi frequency~\cite{Wineland1979,Blatt2009}. For a Franck-Condon transition at the classical turning point this equation reduces to:
\begin{equation}\label{eq:rabi}
\Omega^2 \approx \left|\Omega_0 \int \mathcal{Z}_{n_z'}(r\ped{c}, z) \exp{i k\ped{c} z}\mathcal{Z}_{n_z}(r\ped{c}, z) dz \right|^2.
\end{equation}
The Rabi frequencies therefore depend on $r\ped{c}$ and can be calculated from the longitudinal Born--Oppenheimer wavefunctions $\mathcal{Z}_{n_z}(r,z)$. 

The limits of integration in Eq.~\ref{eq:vband} go from $E\ped{min} = U_{n_z}(0)$ to $E\ped{max} = -h\delta$, an approximation that is valid for a deep vertical lattice where we can neglect tunnelling and the Wannier-Stark ladder~\citep{Lemonde2005}.
The maximum limit corresponds to $U_{n_z+1}(r\ped{c}) < 0$. Atoms with energy $E > -h\delta$ are lost from the trap.
Similar calculations can be carried out for the red sideband, but where the integration limit is then $E\ped{max} = 0$.

With all quantities entering Eq.~\ref{eq:vband} now specified, the sideband spectrum can be evaluated numerically using either the perturbed harmonic oscillator potential or the full lattice potential. Our numerical implementation is available as open-source Python code~\cite{vbandGitHub}. Appendix~\ref{sec:numerical} provides further details on the numerical optimizations used for efficient fitting of experimental spectra.

\subsection{Doppler spectroscopy}

%\sout{An alternative method to derive the temperature of the atoms trapped in the lattice is to probe the atoms with a clock laser beam orthogonal to the lattice direction. In this case, only the radial temperature can be extrapolated by analysing the spectrum with a Gaussian curve. The radial temperature is retrived as}

%\textcolor{orange}{An alternative method for determining the radial atomic temperature is transverse Doppler spectroscopy. In this approach, the clock transition is probed along a weakly confined radial direction, orthogonal to the lattice axis. Assuming that the atomic motion along the probe direction is approximately free and follows a Maxwell–Boltzmann velocity distribution, the Doppler-broadened spectrum has a gaussian profile. The radial temperature can therefore be extracted by fitting the spectrum with a gaussian function and using} $T_r = \frac{m c^2}{k\ped{B} \nu_{0}^2}\sigma^2$, where $\sigma^2$ is the variance of the gaussian profile, $m$ is the atomic mass, $\nu_{0}$ its atomic transition frequency, and c is the speed of light in vacuum~\cite{Gianfrani2016, Wineland1979}.

Another method for determining the atomic temperature is Doppler spectroscopy~\cite{Gianfrani2016}.
The spectrum of a weakly confined atom approaches that of a free atom, exhibiting gaussian Doppler broadening~\cite{Wineland1979}. In an optical lattice, the confinement in the radial direction is weak, so a gaussian profile can approximate atomic spectra recorded along a transverse axis. The radial temperature can be extracted by fitting spectra with a gaussian function and using $T_r = \frac{m c^2}{k\ped{B} \nu_{0}^2}\sigma^2$, where $\sigma^2$ is the variance of the gaussian profile, $m$ is the atomic mass, $\nu_{0}$ its atomic transition frequency, and c is the speed of light in vacuum.

\subsection{Comparison of the models}\label{sec:comparison}
\begin{figure}[h]
	\centering
    \begin{overpic}[width=0.6\linewidth]{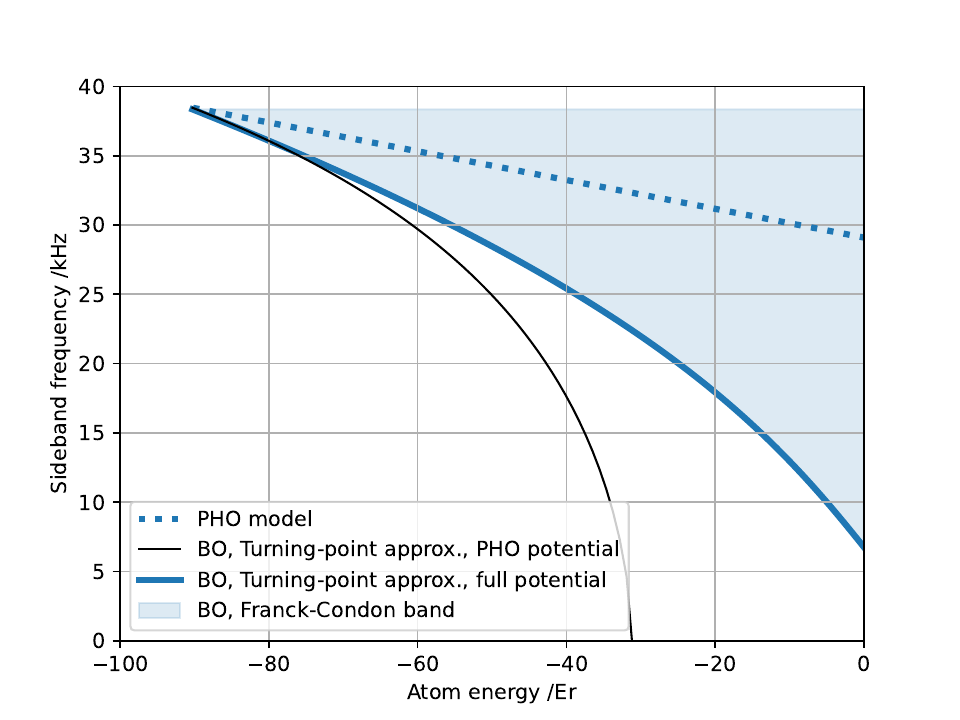}
    \put(0,60){\text{a)}} 
    \end{overpic}
    \begin{overpic}[width=0.6\linewidth]{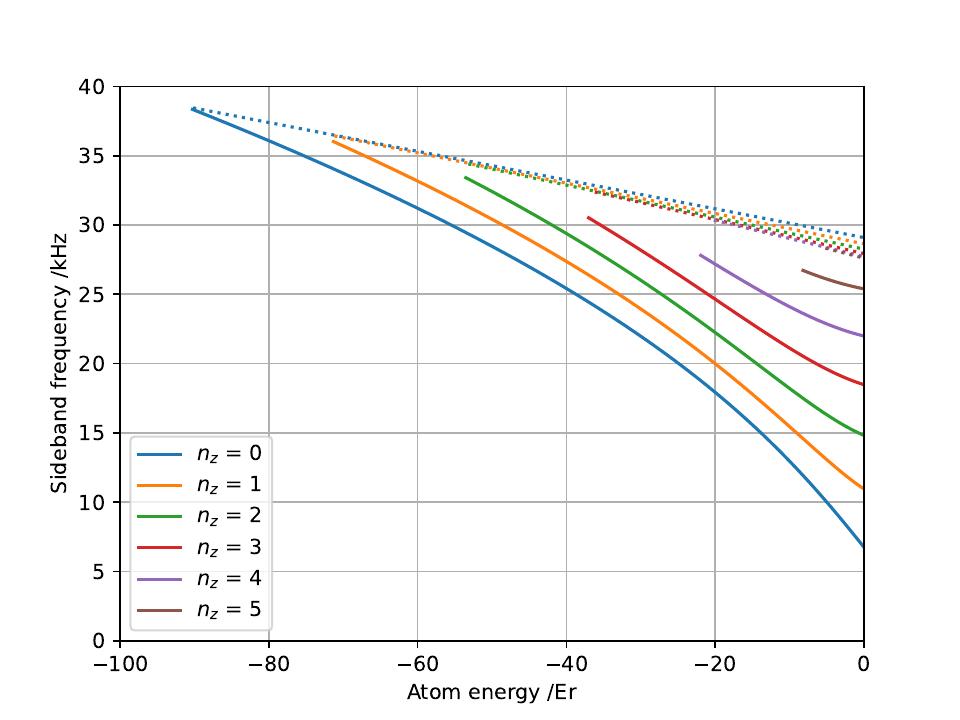}
    \put(0,60){\text{b)}}
    \end{overpic}
    \caption{Comparison of the expected sideband frequency for $^{171}$Yb atoms with energy $E$ in a trap with depth $D = 100\,E\ped{R} \simeq \SI{10}{\micro\kelvin} \cdot k_B$ in the perturbed harmonic oscillator (PHO) and Born-Oppenheimer (BO) models. (a) Detailed view of the blue sideband ($n_z=0\to1$) transition, showing the perturbed harmonic oscillator prediction (dotted blue line), and the Born-Oppenheimer, Franck-Condon, and turning-point approximations applied to the perturbed harmonic oscillator potential (black line) and to the full lattice potential (solid blue line). For the full potential, the Franck-Condon-allowed band is indicated by the light blue shaded region. (b) Expected blue sideband frequency for an atom at energy $E$ for different initial $n_z$. Dotted lines correspond to the perturbed harmonic oscillator approximation (Eq.~\ref{eq:ho-freq}), while the solid line corresponds to the turning-point approximation in the Born-Oppenheimer model for the full lattice potential (Eq.~\ref{eq:bo-freq}).}
	\label{fig:vsblatt-energy}
\end{figure}

The perturbed harmonic oscillator model and the Born-Oppenheimer model introduce different approximations, which result in different functions mapping sideband frequency to atomic energy. However, both models make the simplifying assumption that, at a given laser frequency, for each initial \(n_z\) only atoms with a given energy contribute to the sideband signal. To illustrate the resultant discrepancy, Fig.~\ref{fig:vsblatt-energy} compares the expected sideband frequency as a function of the atomic energy according to the perturbed harmonic oscillator model and the Born-Oppenheimer model applied to either the perturbed harmonic oscillator potential or the full lattice potential. Because the temperature is extracted by fitting the sideband lineshape, the different frequency–energy relations lead to different fitted temperature estimates. As shown, the perturbed harmonic oscillator prediction of the sideband frequency lies near the center of the Franck-Condon-allowed band, whereas the turning-point approximation assumes the most probable transition to occur at the inner edge of the band for both underlying potentials. The predictions of the different models diverge even at low atomic energies, near the bottom of the trap, implying that they will not yield the same temperature estimates. More specifically, the slopes of the two frequency-versus-energy relations differ by approximately a factor of two because of the simplifying assumptions used by each model to select the most probable transition. By contrast, applying the turning-point approximation to the perturbed harmonic oscillator potential or the full lattice potential
results in similar predictions for low-energy atoms and a divergence for hotter atoms, for which the perturbed harmonic oscillator approximation begins to break down.

\begin{figure}[h]
	\centering
	\includegraphics[trim=20 0 0 0, clip, width=0.85\linewidth]{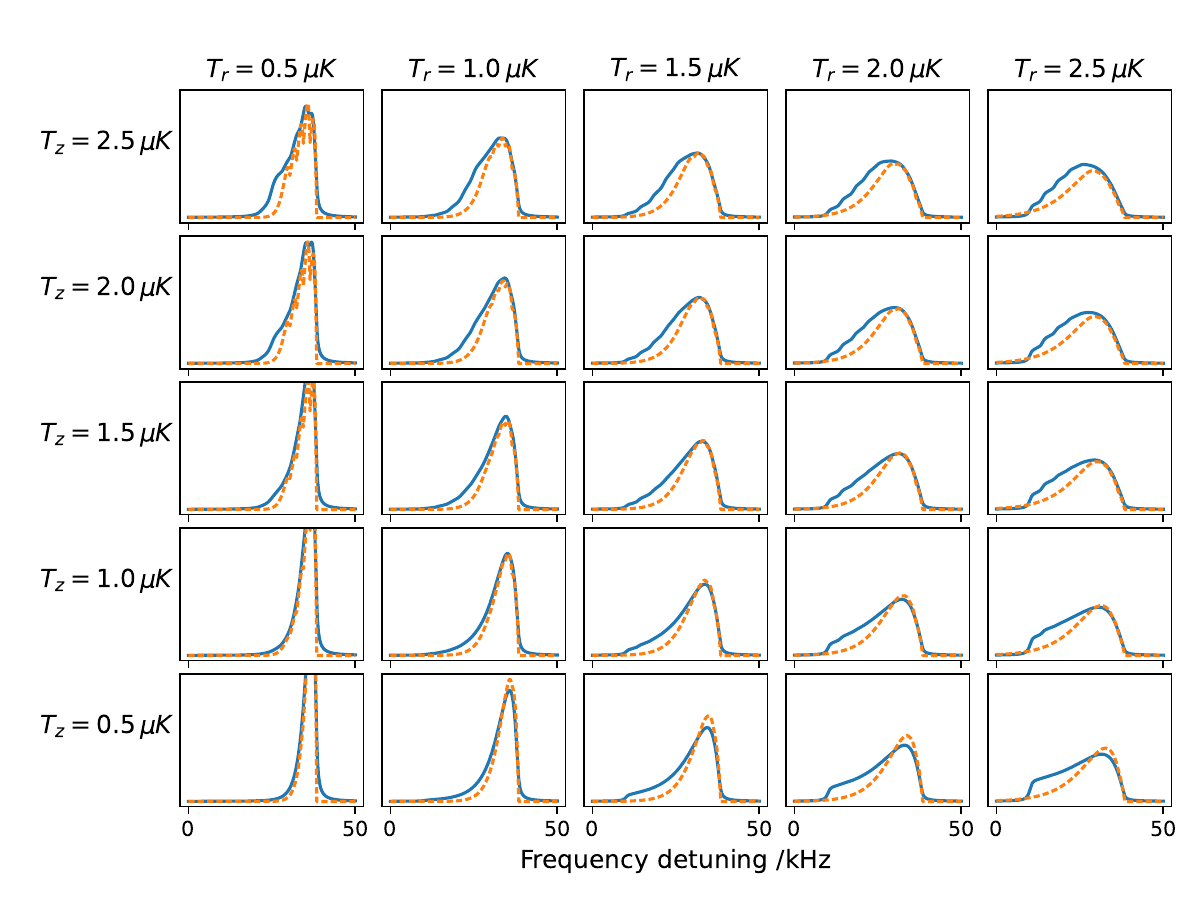}
	\caption{Calculated blue sideband shapes (excitation probability) for Yb atoms in a lattice of depth $D = 100\,E\ped{R}$ for different $T_z$ and $T_r$ in the Born-Oppenheimer model with the full lattice potential (blue solid lines) and perturbed harmonic oscillator model (orange dashed lines). The curves for the perturbed harmonic oscillator model are calculated for twice the stated temperatures (see main text for details).}
	\label{fig:grid100}
\end{figure}

Figure~\ref{fig:grid100} shows the sideband shapes predicted by the two models for Yb atoms in a lattice of depth $D = 100\,E\ped{R}$, at different longitudinal and radial temperatures. Both models yield sideband shapes with a different dependence on $T_z$ and $T_r$. At higher temperatures, Equation~\ref{eq:vband} predicts deviations from the exponential behavior expected from the perturbed harmonic oscillator approximation. Fig.~\ref{fig:grid100} shows sidebands calculated with the perturbed harmonic oscillator model using twice the stated temperature. This empirical factor is used to account for the different slopes between the sideband frequency and the atomic energy shown in Fig.~\ref{fig:vsblatt-energy}.

\subsection{Limitations of the models}\label{sec:limitations}
The perturbed harmonic oscillator model has two main limitations. The first is the use of harmonic oscillator selection rules, whose effects remain apparent even near the bottom of the trap, where the harmonic approximation should work well. As discussed in Sec.~\ref{sec:comparison}, the perturbed harmonic oscillator and the turning-point predictions differ in this regime even when the same underlying potential is used (black and blue dotted lines in Fig.~\ref{fig:vsblatt-energy}). Our comparison suggests that harmonic-oscillator selection rules do not fully capture the transition structure of atoms in the lattice potential, for which the Born–Oppenheimer treatment provides a more general description. The anharmonicity of the trap is expected to relax the harmonic oscillator selection rules, allowing transitions that are forbidden in the harmonic approximation and resulting in a band analogous to the Franck-Condon case. 
%The treatment of the matrix elements for these transitions will result in a dominant contribution toward the inner edge of the band, as assumed in the turning-point approximation, thereby reducing the inferred temperature.
The second limitation is the breakdown for sufficiently hot atoms. We estimate that, for $T_z=T_r\gtrsim0.2\,D/k\ped{B}$, most of the atomic population occupies states far from the trap minimum, outside the regime of validity of the approximation. In this case, agreement between the perturbed harmonic oscillator model and the experimental data may be improved by extending the calculation to states with energies $E>0$ or by including longitudinal states that are not bound at the considered lattice depth. Although these extensions can improve agreement with the measured spectra, they may lead to an overestimate of the extracted atomic temperature.
%The perturbed harmonic oscillator model has two main limitations: the breakdown of the harmonic approximation at sufficiently high temperatures and the use of harmonic oscillator selection rules.
%For sufficiently hot atoms, the lattice potential can no longer be accurately described by the second order harmonic expansion.
%The anharmonicity of the trap is expected to relax the harmonic oscillator selection rules, allowing transitions that are forbidden in the harmonic approximation.
%Moreover, we estimated that when the atomic temperature $T_z = T_r \gtrsim 0.2 D/k_\text{B}$ a significant fraction of atoms populates states with energies far from the trap minimum.
%In this case, agreement between this model and the experimental data may be improved by including states with energies exceeding the lattice trap depth ($E > 0$).
%As a result, the perturbed harmonic oscillator model is expected to overestimate the atomic temperature.

The Born-Oppenheimer approximation and the Franck-Condon principle rely on the separation of timescales between fast and slow degrees of freedom. This separation is typically well justified in 1D optical lattices owing to the much stronger longitudinal confinement: in our case, $\nu_z / \nu_r \approx 400$. Yet, the turning-point approximation assigns each sideband frequency to the single inner edge of the band of Franck-Condon-allowed transitions, instead of the full band. This is likely to lead to an underestimate of the atomic temperature.

Sideband spectra are also used to extract the longitudinal trap frequency $\nu_z$ from the outer sideband edge, and thus determine the lattice depth.
Both the perturbed harmonic oscillator and the Born-Oppenheimer models predict a sharp outer edge (ignoring the broadening of the underlying Lorentzian) from atoms at the bottom of the trap where the density of states is zero. However, the edge frequency lies within the Franck-Condon-allowed band for all atoms, so additional signals not accounted for by either model may introduce systematic effects in the determination of the trap depth.

A more complete treatment will require calculating the Franck-Condon factors for all relevant combinations of initial and final states.
This would account for the full band of allowed transitions, rather than assigning each sideband frequency to a single atomic population, but would also require treating a complex and highly degenerate dynamics (e.g., \cite{Bohn1999, Morris1983}), including the effects of the high-intensity probe laser.

Finally, transverse Doppler spectroscopy assumes that the atomic motion can be treated as approximately free, since the atoms are probed along a weakly confined direction~\cite{Wineland1979}.
It therefore neglects coupling between the longitudinal and radial motions, as well as other features of the trap.

\section{Experimental setup}
The models described above were tested on IT-Yb1, the ytterbium optical lattice clock developed at INRIM. The experimental setup has been detailed in previous works~\cite{Pizzocaro2020, Goti2023}. 
In IT-Yb1, the atoms are trapped in a one-dimensional vertical lattice operating at the magic wavelength after cooling in a two-stage magneto-optical trap. The maximum lattice power delivered to the atoms is $\SI{2.5}{W}$, with a beam waist of $\SI{62}{\micro m}$. This configuration allows us to perform sideband spectroscopy across a range of depths of the magic trap, from approximately $D = 50\,E\ped{R}$ to $D = 700\,E\ped{R}$, corresponding to trap depths of approximately $\sim$ \SI{5}{\micro K} to $\sim$ \SI{70}{\micro K}, longitudinal trap frequencies between$\sim$\SI{30}{kHz} and$\sim$\SI{110}{kHz}, and radial trap frequencies between$\sim$\SI{80}{Hz} and$\sim$\SI{300}{Hz}. The experimental data for sideband spectroscopy were acquired using a clock pulse propagating along the lattice axis, with a duration of \SI{4}{ms} and an optical power of approximately \SI{1.5}{mW}, resulting in a carrier linewidth of approximately \SI{2}{kHz}. Alternatively, to probe the radial atomic motion by means of Doppler spectroscopy, the clock transition was interrogated with a collimated beam propagating orthogonally to the lattice axis. The interrogation pulse duration was again \SI{4}{ms}, while the optical power delivered to the atoms was approximately \SI{220}{\micro W}. The resulting spectra were fitted with Gaussian profiles, from which the Doppler width was extracted and converted into a radial temperature. In all measurements, the lattice beam remained continuously on at a constant power. The lattice power was actively stabilized via a photodiode-based feedback loop, ensuring a reproducible trap depth and enabling a direct comparison between the two techniques.

In our experiment, post-cooling is implemented using clock-line-mediated Sisyphus cooling, a technique recently demonstrated by Chen et al. at NIST~\cite{Chen2024} to reduce the atomic longitudinal and radial temperatures. During Sisyphus cooling, the atoms are illuminated with two perpendicular 1D $\SI{1389}{nm}$ optical lattices, propagating in the transverse direction of the magic wavelength lattice. At the same time, the clock transition is stimulated with a resonant pulse at $\SI{578}{nm}$.
Each $\SI{1389}{nm}$ lattice is generated by retro-reflecting a beam of approximately $\SI{200}{\micro W}$ with linear polarization, the same for both axes. The $\SI{1389}{nm}$ laser is blue-detuned by about $\SI{80}{MHz}$ from the $^3\text{P}_0$ $\rightarrow$ $^3\text{D}_1$ transition. For simplicity, this detuning is kept constant throughout the entire clock cycle, as the repumping process at the end of the cycle remains effective under these conditions.
The $\SI{578}{nm}$ light, with a Rabi frequency of approximately $\SI{5}{kHz}$, is sent towards the atoms using the same beam path used for sideband spectroscopy. The Sisyphus cooling sequence begins during the final \SI{30}{ms} of the magneto-optical trap stage and continues after for a variable duration $T_{\mathrm{cool}}$ with atoms trapped in the magic lattice. The post-cooling stage is followed by state preparation and clock spectroscopy. 
This configuration enables precise control of atomic temperature at various trap depths, enabling experimental validation of the theoretical models described in the previous section.
Hereafter, the term ``Born-Oppenheimer model" refers to the implementation of Eq.~\ref{eq:vband} using the full lattice potential.

\section{Results}

\begin{figure}[h!]
  \centering
  \begin{overpic}[width=0.65\linewidth]{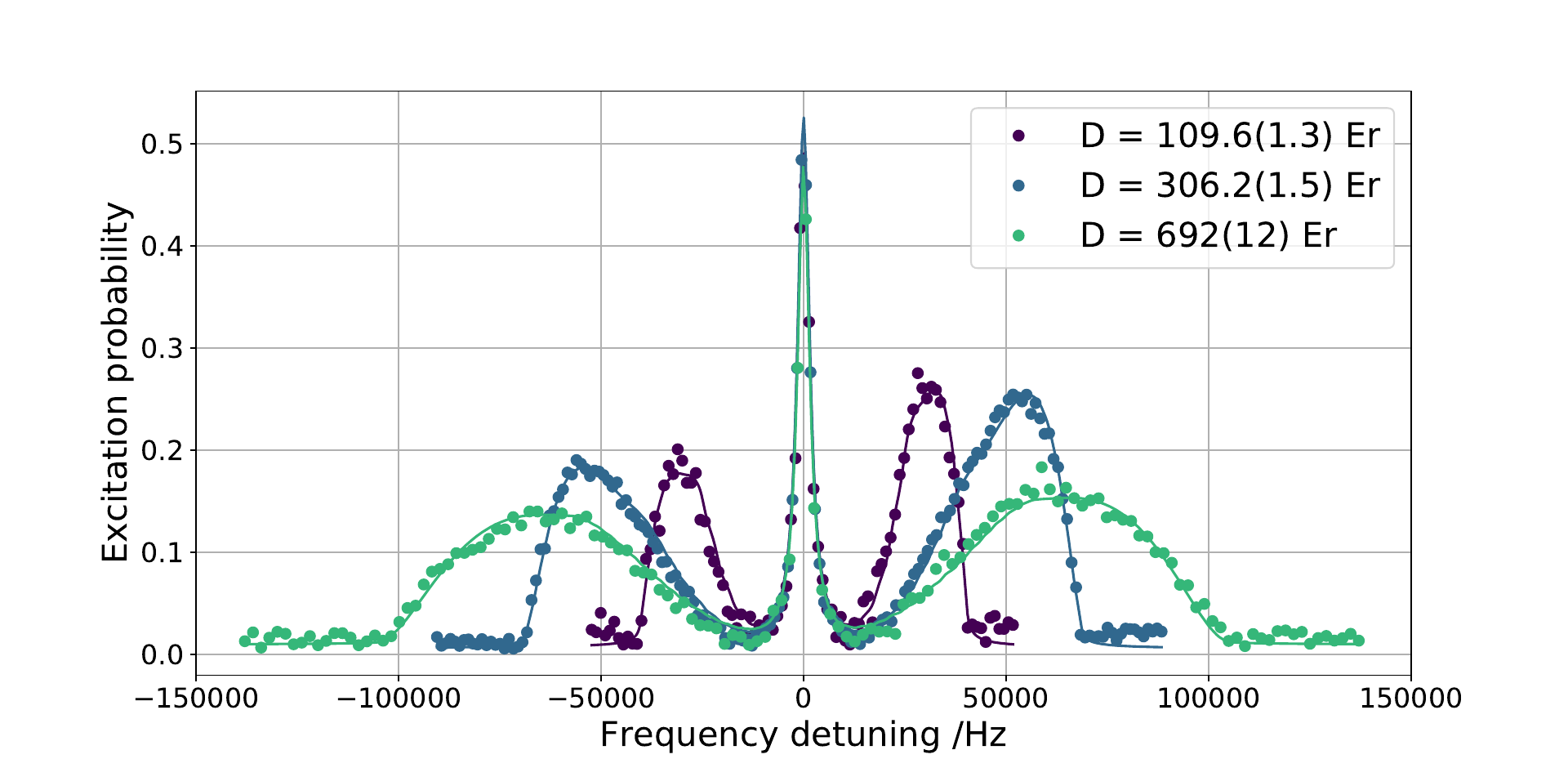}
    \put(4,40){\text{a)}} % Cambia i valori per posizionare meglio
  \end{overpic}
  % Pannello (b)
  \begin{overpic}[width=0.65\linewidth]{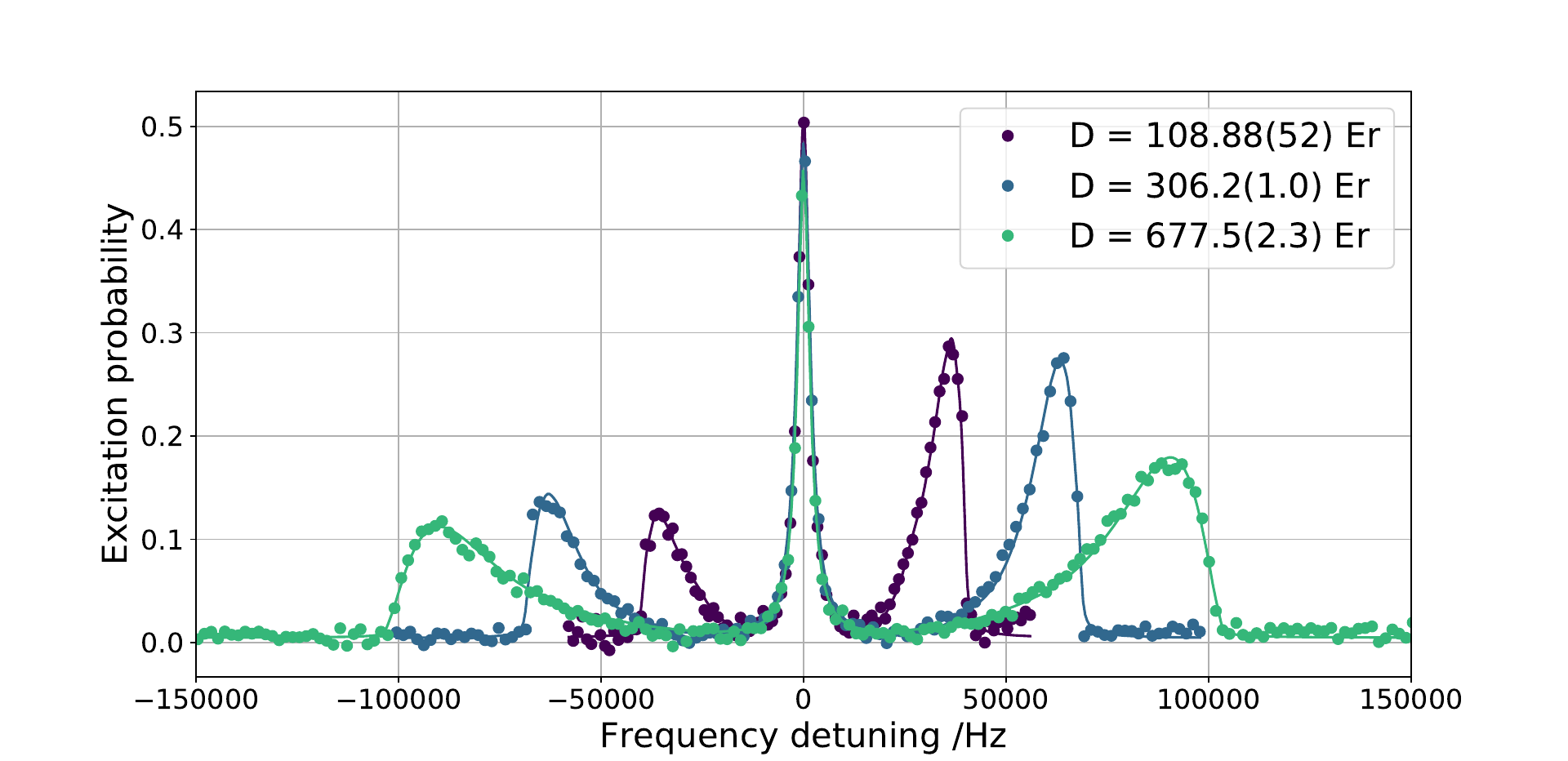}
    \put(4,40){\text{b)}}
  \end{overpic}
  % Pannello (c)
  \begin{overpic}[width=0.65\linewidth]{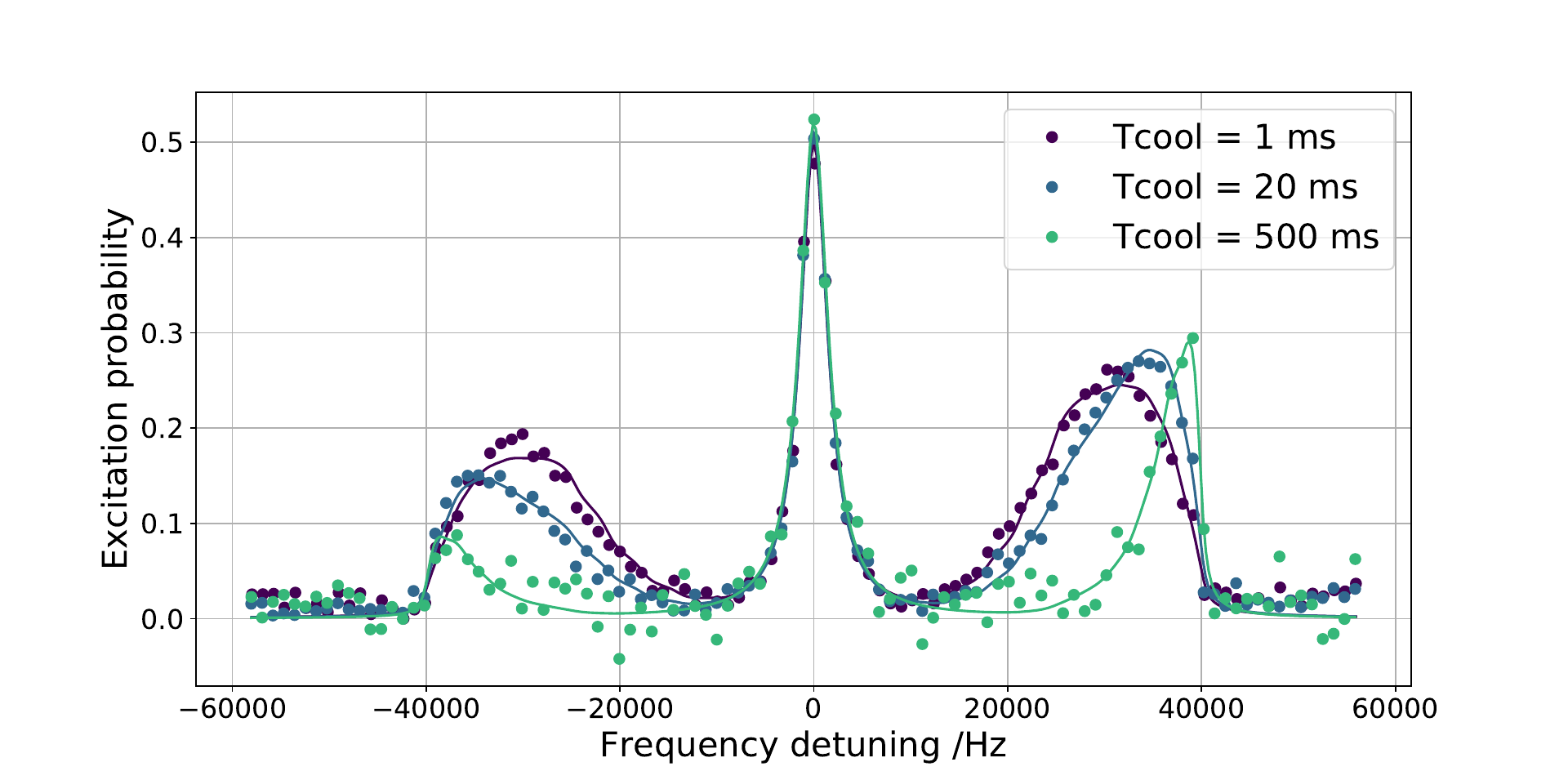}
    \put(4,40){\text{c)}}
  \end{overpic}
  \caption{Sideband spectra collected at different trapping conditions fitted with the Born-Oppenheimer model: a) No post-cooling in the loading sequence; b) Sisyphus cooling with $T_{\mathrm{cool}} = \SI{50}{ms}$. In both panels, purple, blue and green dots represent data collected at about $D = 100\,E\ped{R}$, $D = 300\,E\ped{R}$ and $D = 700\,E\ped{R}$, respectively. c) Sideband spectra obtained for a lattice depth of $D = 100\,E\ped{R}$ and for different $T_{\mathrm{cool}}$: purple, blue and green dots represent data collected by applying a $T_{\mathrm{cool}}$ of $\SI{1}{ms}$, $\SI{20}{ms}$ and $\SI{500}{ms}$, respectively. Solid lines represent fits obtained using Eq.~\ref{eq:vband} ~\cite{vbandGitHub}.}
  \label{fig:Sideband}
\end{figure}

Sideband spectra obtained with IT-Yb1 are presented in Fig.~\ref{fig:Sideband}.
Panel~a) shows experimental data without post-cooling, panel~b) shows data with Sisyphus cooling implemented for $T_{\mathrm{cool}}=\SI{50}{ms}$, and panel~c) shows data for varying cooling durations $T_{\mathrm{cool}}$. Different colors in panels a) and b) correspond to spectra taken at different lattice depths $D$, while in panel c) they represent spectra for different values of $T_{\mathrm{cool}}$. 
Each spectrum is fitted using models based on the Born-Oppenheimer approximation (Eq.~\ref{eq:vband}), displayed as solid lines, and the perturbed harmonic oscillator approximation (Eq.~\ref{eq:blatt-blue}, not shown). The fit results are in good agreement with the experimental data across a wide range of trapping conditions. From the fits, the associated lattice depth ($D$), longitudinal temperature ( $T_z$), and radial temperature ($T_r$) are obtained. 

\begin{figure}[h!]
	\centering
        \includegraphics[trim=5 40 0 70, clip, width=0.7\linewidth]{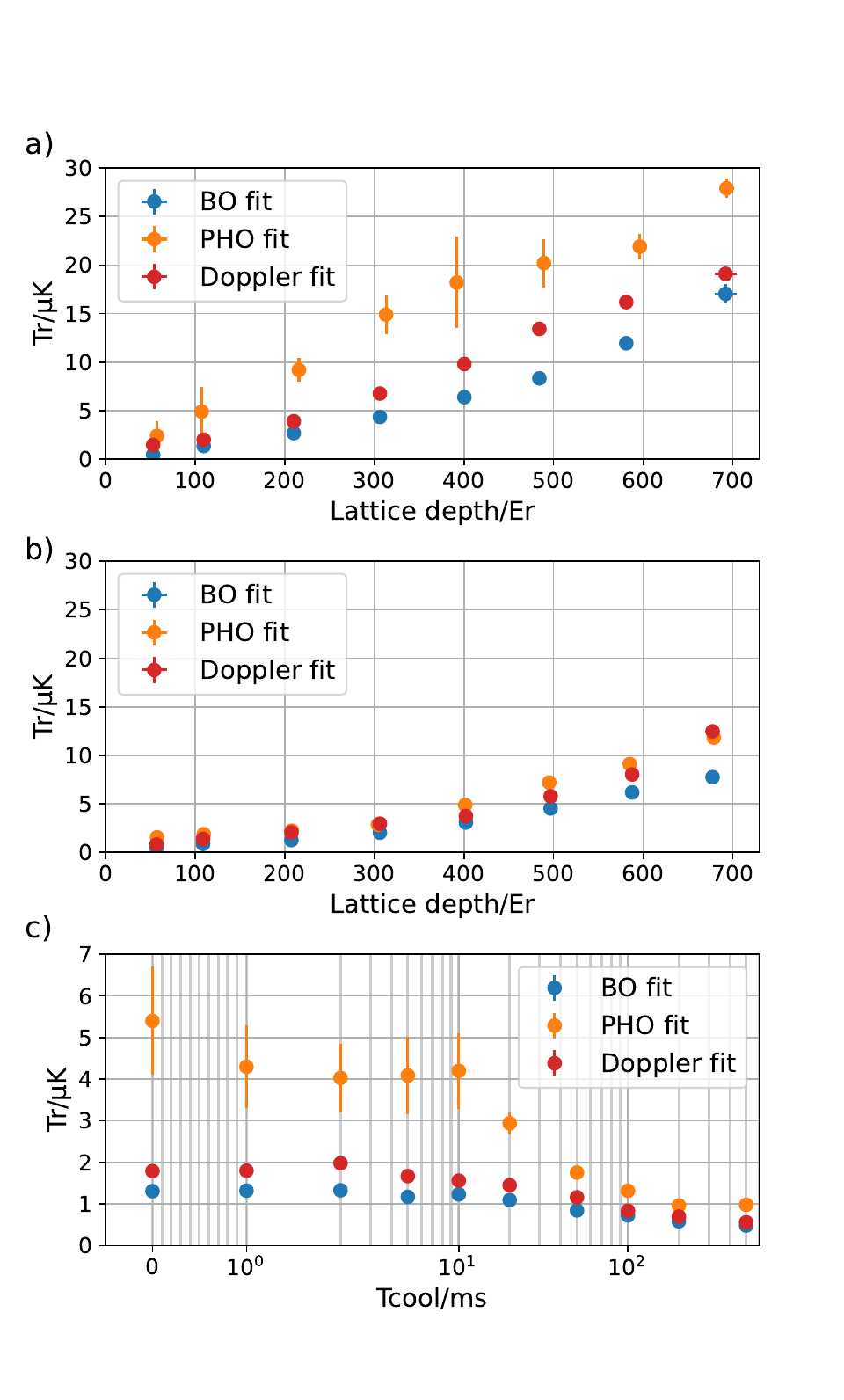}
	\caption{Radial atomic temperatures as a function of lattice depth: a) without post-cooling; b) with a Sisyphus cooling pulse with $T_{\mathrm{cool}}= \SI{50}{ms}$; c) as a function of $T_{\mathrm{cool}}$, for a lattice depth of $D = 100\,E\ped{R}$. The temperatures shown in this figure are obtained from the analysis of spectra, with representative examples shown in Fig.\ref{fig:Sideband}. Blue and orange dots represent the results obtained by fitting sideband spectra using the Born-Oppenheimer (Eq.~\ref{eq:vband}) and perturbed harmonic oscillator (Eq.~\ref{eq:blatt-blue}) models, respectively. Red dots show radial temperatures derived from Doppler spectroscopy performed under the same trapping conditions. Error bars are derived from the fits: in some cases, they are smaller than the marker size and therefore not visible. For $^{171}$Yb atoms trapped in the magic-wavelength optical lattice, $100\,E\ped{R}/k_B\simeq\SI{10}{\micro\kelvin}$.}
	\label{fig:NoSis}
\end{figure}

\begin{figure}[h!]
	\centering
	\includegraphics[trim=5 40 0 70, clip, width=0.7\linewidth]{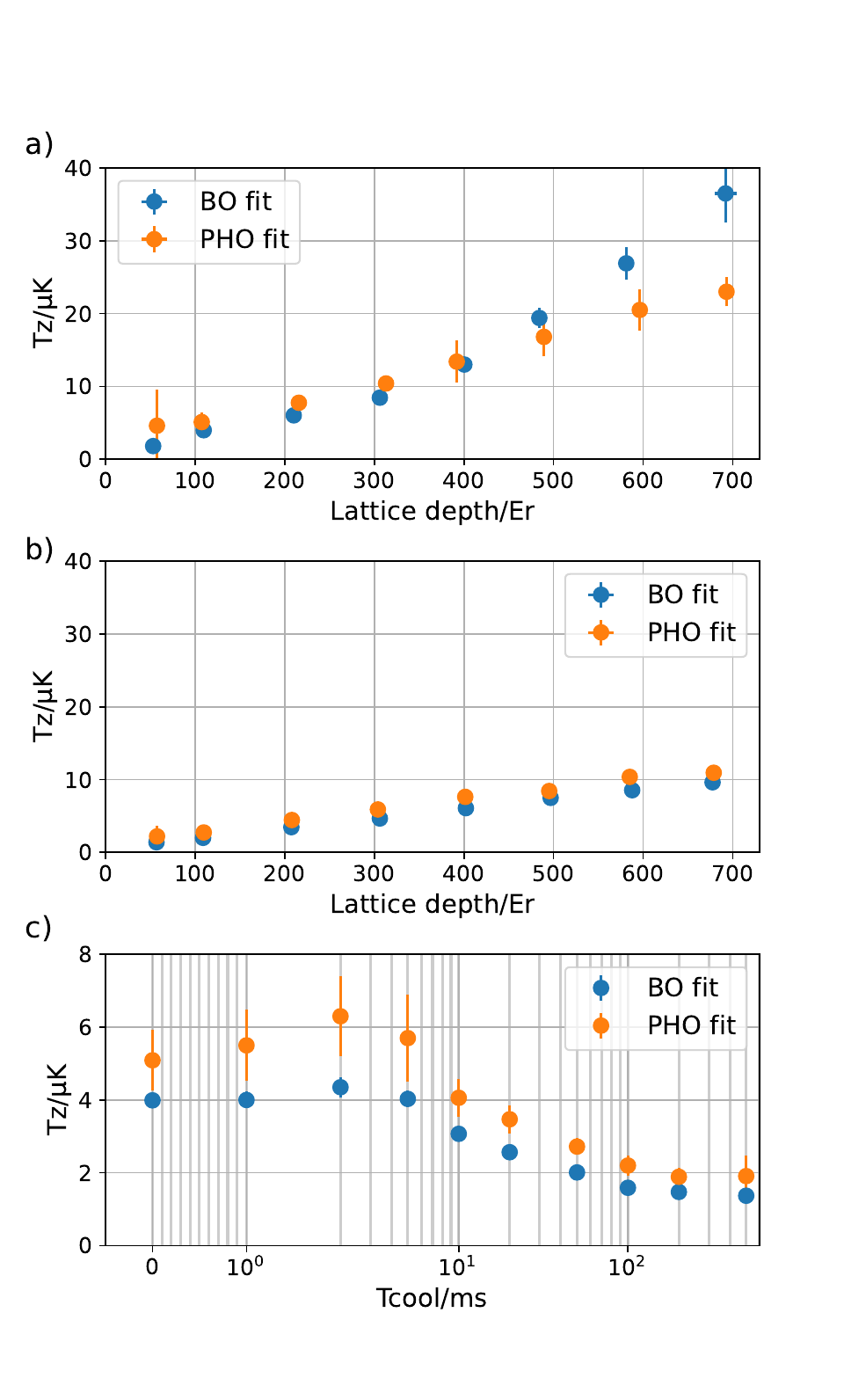}
	\caption{Longitudinal atomic temperatures as a function of lattice depth: a) without post-cooling; b) with a Sisyphus cooling pulse with $T_{\mathrm{cool}}= \SI{50}{ms}$ in the loading sequence; c) as a function of $T_{\mathrm{cool}}$, for a lattice depth of $D = 100\,E_R$. The temperatures shown in this figure are obtained from the analysis of spectra, with representative examples shown in Fig.\ref{fig:Sideband}.
    Blue and orange dots represent the results obtained by fitting sideband spectra using the Born-Oppenheimer (Eq.~\ref{eq:vband}) and perturbed harmonic oscillator (Eq.~\ref{eq:blatt-blue}) models, respectively. Error bars are derived from the fits: in some cases, they are smaller than the marker size and therefore not visible. For $^{171}$Yb atoms trapped in the magic-wavelength optical lattice, $100\,E\ped{R}/k_B\simeq\SI{10}{\micro\kelvin}$.} 
	\label{fig:Sis}
\end{figure}

The values of $D$, $T_r$, and $T_z$ obtained from fits to each spectrum are compared in Figs.~\ref{fig:NoSis} and~\ref{fig:Sis}. In analogy with Fig.~\ref{fig:Sideband}, panel~a) in both figures shows results without post-cooling, panel~b) shows results with Sisyphus cooling with $T_{\mathrm{cool}}=\SI{50}{ms}$, and panel~c) shows results for varying Sisyphus-cooling durations. The orange and blue dots represent the temperatures obtained using Eq.~\ref{eq:blatt-blue} and Eq.~\ref{eq:vband}, respectively. In addition, the radial temperatures estimated via Doppler spectroscopy are depicted as red dots. Since the lattice depth cannot be directly extracted from Doppler spectroscopy, in Fig.~\ref{fig:NoSis} the depth associated with the Doppler data is taken from sideband spectroscopy measurements performed under the same trapping conditions and analyzed using the Born-Oppenheimer model.

A comparison between panels a) and b) of Fig.~\ref{fig:NoSis} and~\ref{fig:Sis} shows that the application of a Sisyphus pulse with $T_{\mathrm{cool}} = \SI{50}{ms}$ results in a twofold reduction in both longitudinal and radial atomic temperatures across all investigated lattice depths, compared with the corresponding measurements performed without post-cooling. Under typical trapping conditions in IT-Yb1, with a lattice depth of $D = 100\,E\ped{R}$ and $T_{\mathrm{cool}} = \SI{50}{ms}$, the temperatures decrease to $T_z = \SI{2}{\micro K}$ and $T_r = \SI{0.8}{\micro K}$. For longer Sisyphus pulses, the reduction in longitudinal temperature becomes even more significant (see panel c) of Fig.~\ref{fig:Sis}).
Moreover, Sisyphus cooling increases the number of atoms trapped in the magic lattice by a factor of two for depths below $100\,E\ped{R}$, enabling operation of IT-Yb1 at a reduced working depth of $50\,E\ped{R}$. This lower depth is advantageous for mitigating lattice light shift and collisional shifts.

\begin{figure}[h!]
	\centering
	\includegraphics[trim= 60 10 0 40, clip, width=0.8\linewidth]{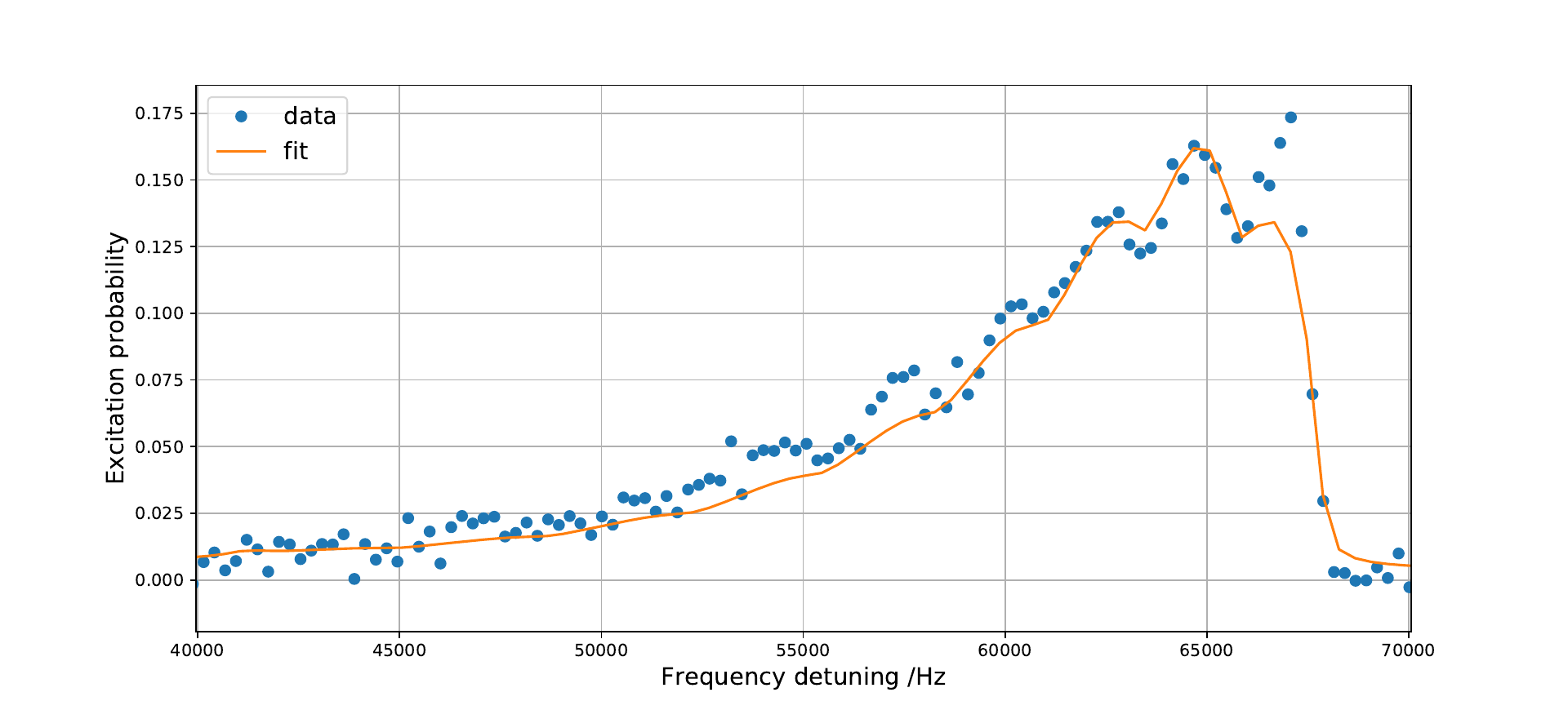}
	\caption{Blue sideband spectrum acquired with IT-Yb1 at a lattice depth of $D = 300\,E\ped{R}$, using Sisyphus cooling with $T_{\mathrm{cool}}=\SI{200}{ms}$. Blue dots represent the experimental data, while the solid orange line is a fit based on the Born-Oppenheimer model. The fit yields atomic temperatures of $T_z = \SI{4.90(10)}{\micro K}$ and $T_r = \SI{0.99(3)}{\micro K}$. Under these trapping conditions, transitions from different longitudinal vibrational states $n_z$ are spectrally resolved, from $n_z=0$ on the right to progressively higher $n_z$ states toward the left.}
	\label{fig:blueSideband}
\end{figure}

Figure~\ref{fig:Sideband} shows spectra in which the contributions from different longitudinal vibrational states $n_z$ are not resolved. Resolving this structure requires a sufficiently high $T_z$ to populate multiple $n_z$ states and a sufficiently low $T_r$ to reduce broadening~\cite{Zhang2022, Bothwell2025a}. Equation~\ref{eq:vband} can describe spectra in which the individual $n_z$ contributions are resolved. As an example, Fig.~\ref{fig:blueSideband} presents a spectrum recorded with IT-Yb1 in an uncommon configuration that maximizes this effect. The fit yields a lattice depth $D = 300\,E\ped{R}$, and longitudinal and radial temperatures are $T_z = \SI{4.90(10)}{\micro K}$ and $T_r = \SI{0.99(3)}{\micro K}$. The measured spectrum shows a slight deviation from the two-temperature distribution (Eq.~\ref{eq:boltzmann}): the model predicts the $n_z=1$ peak to be the highest, whereas the $n_z=0$ peak is the highest in the experimental data.

\begin{figure}
    \centering
    \includegraphics[width=0.7\linewidth]{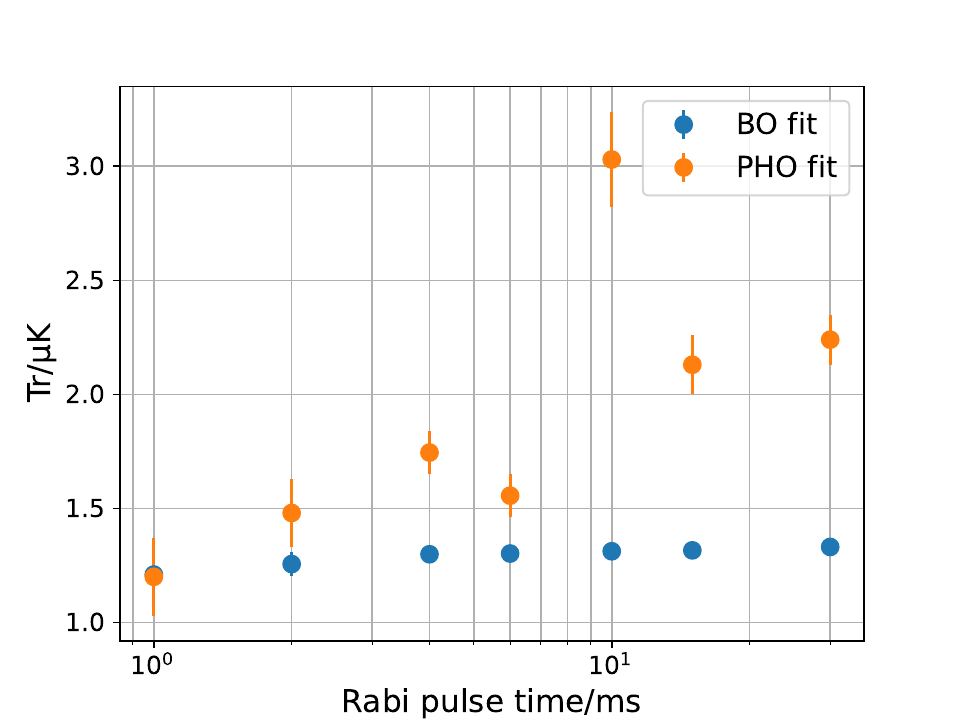}
    \caption{Fitted radial temperature measured as a function of Rabi time. Blue dots show the results obtained with the Born-Oppenheimer model, while orange dots show those obtained with the perturbed harmonic oscillator model.}
    \label{fig:tr_vs_tr}
\end{figure}

Finally, we tested the dependence of the extracted atomic temperature on the spectroscopy pulse duration (Rabi time) over the range from \SI{1}{ms} to \SI{30}{ms}. The results obtained with both models are shown in Fig.~\ref{fig:tr_vs_tr}.
The Born-Oppenheimer model gives consistent radial temperature estimates as the Rabi time is increased, in contrast to the behavior observed with the perturbed harmonic oscillator model~\cite{Blatt2009}.
Longer Rabi times increase the sideband excitation, including along the shallow sideband slope facing the carrier. We hypothesize that the perturbed harmonic oscillator model interprets the additional excitation in this region as a higher radial temperature, because it predicts little sideband contribution in this spectral region.

\section{Discussion}
The estimated atomic temperatures depend on the model used (perturbed harmonic oscillator, Born-Oppenheimer, or Doppler spectroscopy).
The Born-Oppenheimer model consistently yields lower temperature estimates than the perturbed harmonic oscillator approximation. 
This systematic difference can be understood from Fig.~\ref{fig:vsblatt-energy}: for a given energy, the perturbed harmonic oscillator approximation predicts a higher sideband frequency and therefore associates the measured spectrum with a higher temperature.
For the radial temperature, the Born–Oppenheimer and perturbed harmonic oscillator estimates differ by up to a factor of four in the absence of post-cooling (Fig.~\ref{fig:NoSis}a). When clock-line-mediated Sisyphus cooling is included in the loading sequence, this difference is reduced to approximately a factor of two (panels b and c of Fig.~\ref{fig:NoSis}), consistent with prediction of Fig.~\ref{fig:vsblatt-energy} for atoms at the bottom of the trap. The longitudinal temperature $T_z$ exhibits smaller differences of up to a factor of 1.4 with Sisyphus-cooling applied (Fig.~\ref{fig:Sis}b and c).

%The relative slopes of the two predictions are consistent with the twofold difference observed between the radial temperature estimates for the Sisyphus-cooled spectra.

%In particular, in the absence of post-cooling, the Born-Oppenheimer and perturbed harmonic oscillator approximations yield values that may differ by a factor of 2 for both radial and longitudinal temperatures (see panels a of Fig.~\ref{fig:NoSis} and~\ref{fig:Sis}).
%This discrepancy is reduced when clock-line-mediated Sisyphus cooling is included in the loading sequence (see panels b of Fig.~\ref{fig:NoSis} and~\ref{fig:Sis}) and further decreases with $T_{\mathrm{cool}}$ (see panels c of Fig.~\ref{fig:NoSis} and~\ref{fig:Sis}).

%The Born-Oppenheimer model consistently yields slightly lower temperature estimates than the perturbed harmonic oscillator approximation. This difference in the atomic temperature can be explained by looking at Fig.~\ref{fig:vsblatt-energy} that compares the expected sideband frequency for atoms with a certain energy in the trap. The perturbed harmonic oscillator approximation will predict a higher sideband frequency for atoms with a given energy, so it will result in higher measured temperatures.

Since the lattice shift depends on the temperature of the atoms trapped in the lattice and their motional state~\cite{Katori2015, Ovsiannikov2016}, it is important to assess how potential discrepancies between different models impact the estimation of this shift.
To evaluate the lattice light shift, we use the model developed by Beloy et al.~\cite{Beloy2020}:
\begin{equation}
\frac{\delta\nu}{\nu}= - D {\Delta\tilde{\alpha}_{E1}} X - D \Delta\tilde{\alpha}_{M1E2} Y - D^2 \Delta\tilde{\beta} Z
\label{eq:shift}
\end{equation}
where $\Delta\tilde{\alpha}_{E1}$, $\Delta\tilde{\alpha}_{M1E2}$ and $\Delta\tilde{\beta}$ are the differential polarizability, multipolarizability, and hyperpolarizability between the ground ($^1\text{S}_0$) and excited ($^3\text{P}_0$) states, respectively, expressed in units of $\hbar\nu$. In the present analysis, we use $\Delta\tilde{\alpha}_{E1}=0$ (magic wavelength), $\Delta\tilde{\alpha}_{M1E2}\simeq -1.4\times10^{-18} /E\ped{R}$, and $\Delta\tilde{\beta}=\num{-2e-21} /E\ped{R}^2$, consistent with those reported in Refs.~\cite{Bothwell2025a, Nemitz2019}. 

The coefficients $X$, $Y$, and $Z$ are dimensionless fractional depths that convert the nominal trap depth $D$ into the effective lattice potential depths experienced by the atoms.
They can be evaluated numerically using either the Born-Oppenheimer model of Beloy et al.~\cite{Beloy2020} or the perturbed harmonic oscillator model by Ushijima et al.~\cite{Ushijima2018}, here implemented using the analytical formulation called ``modified Ushijima model'' in Ref.~\cite{Beloy2020}.
Because these coefficients depend on the motional-state distribution, they are sensitive to the atomic temperature used in the calculation.
As a result, the temperatures extracted with the perturbed harmonic oscillator, Born-Oppenheimer, and Doppler analyses lead to different estimates of effective depths and of the lattice light shift.

\begin{figure}[h!]
	\centering
        \includegraphics[trim=0 40 0 58, clip, width=0.7\linewidth]{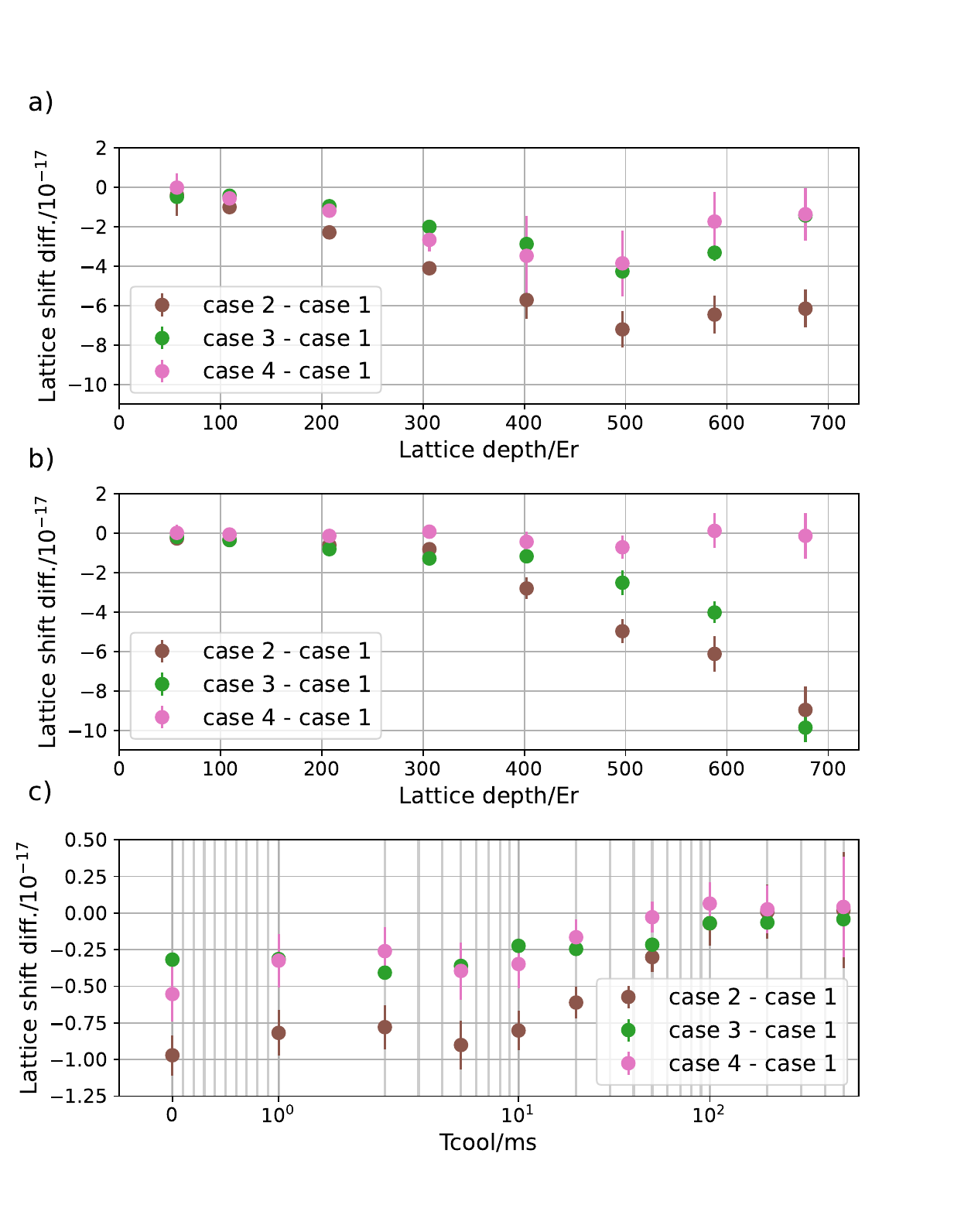}
	\caption{Lattice shift differences obtained using different models for sideband spectrum analysis and shift calculation. Case 1, 2, 3 and 4 refer to simulation parameters reported in Table~\ref{tab:cases}. Panels a), b) and c) correspond to measurements performed without post-cooling, with Sisyphus cooling at $T_{\mathrm{cool}}=\SI{50}{ms}$, and with variable $T_{\mathrm{cool}}$, respectively.}
	\label{fig:Tcool}
\end{figure}

Table~\ref{tab:cases} summarizes the different combinations of lattice shift models and temperature fits we have considered in this work.
The results of this simulation are presented in Fig.~\ref{fig:Tcool}, where panels a), b) and c) correspond to the experimental cases with no post-cooling, with a Sisyphus cooling pulse with $T_{\mathrm{cool}}=\SI{50}{ms}$, and a variable duration of $T_{\mathrm{cool}}$, respectively. The brown, green and pink dots represent the lattice shift differences obtained by subtracting the estimated shift values for Case 1 and those for Case 2, Case 3 and Case 4, respectively. In Case 3, the radial temperature $T_r$ was derived from Doppler spectroscopy, while the longitudinal temperature $T_z$ was obtained using the Born-Oppenheimer model applied to sideband spectra recorded under the same experimental conditions. As a consequence of this modelling choice, the uncertainties associated with the green dots are smaller than those associated with lattice shift differences between models that use the perturbed harmonic oscillator and Born-Oppenheimer fit.
\begin{table}[h!]
\centering
\begin{tabular}{|c|c|c|}
\hline
 & Lattice shift model & Temperature fit \\
\hline
Case 1 & Beloy et al.~\cite{Beloy2020} & $T_z$ = BO fit, $T_r$ = BO fit\\
Case 2 & Beloy et al.~\cite{Beloy2020} & $T_z$ = PHO fit, $T_r$ = PHO fit\\
Case 3& Beloy et al.~\cite{Beloy2020} & $T_z$ = BO fit, $T_r$ = Doppler fit \\
Case 4 & Ushijima et al.~\cite{Ushijima2018, Beloy2020} & $T_z$ = PHO fit, $T_r$ = PHO fit \\
\hline
\end{tabular}
\caption{Lattice shift simulations parameters in terms of model utilized and fit temperatures employed.}
\label{tab:cases} 
\end{table}

In the absence of post-cooling, the lattice light shift shows a significant dependence on the model used to estimate the atomic temperature. Specifically, shifts evaluated with Case 1 and Case 2 differ at the level of \num{1e-17}, even at relatively shallow lattice depths of $D = 100\,E\ped{R}$ (brown dots in Fig~\ref{fig:Tcool}a). When a Sisyphus cooling pulse with $T_{\mathrm{cool}}=\SI{50}{ms}$  is applied, this model dependence becomes less pronounced. In this case, for lattice depths of $D = 100\,E\ped{R}$, the discrepancy in the lattice shift between the fit models is about $4\times10^{-18}$. However, at higher lattice depths, the inconsistency increases, reaching values up to 8$\times10^{-17}$ (see panel b) of Fig~\ref{fig:Tcool}).
Interestingly, the difference in lattice shift decreases when comparing cases with both the temperature fit and the lattice shift model based on the same approximation, such as Case 1 and Case 4, which are based on Born-Oppenheimer and perturbed harmonic oscillator approximations, respectively (see pink dots in Fig.~\ref{fig:Tcool}).

As shown in Fig.~\ref{fig:Tcool}c, reducing the atomic temperature in the lattice by increasing the duration of the Sisyphus cooling stage leads to a progressive decrease in the discrepancies between models. For sufficiently long cooling times $T_{\mathrm{cool}}>\SI{100}{ms}$, i.e. for very cold atoms, the disagreement between different models is reduced to the $1\times10^{-18}$ level, still significant for the best optical clocks.

Finally, we note that the use of the model presented by Brown et al.~\cite{Brown2017} assumes that the atomic temperatures $T_z$ and $T_r$ are linear as a function of the lattice depth $D$. As the extraction of atomic temperatures is dependent on the chosen fit model, this may introduce a model-dependent bias in the evaluation of the lattice frequency shift based on this strategy.
%\textcolor{red}{Finally, we note that Eq.~\ref{eq:shift} can be reduced to the form $aD+bD^2$, where $a$ and $b$ are constant coefficients, if the radial and longitudinal temperatures vary linearly with the lattice depth $D$~\cite{Brown2017,Beloy2020}. However, the temperatures inferred from sideband spectra depend on the thermometry model used and may therefore exhibit different dependences on $D$. Consequently, the validity of this simplification is itself model dependent.}

\section{Conclusions}
In this work, we analyzed the spectra of ytterbium atoms trapped in an optical lattice at depths ranging from $50\,E\ped{R}$ to $700\,E\ped{R}$, and at different atomic temperatures.
The application of clock-line-mediated Sisyphus cooling~\cite{Chen2024} led to a twofold reduction in both radial and longitudinal temperatures and enabled clock operation at lattice depths as low as $50\,E\ped{R}$.

In addition to the perturbed harmonic oscillator and Doppler models traditionally used to determine atomic temperatures, we employed a model~\cite{vbandGitHub} based on the Born-Oppenheimer approximation (Eq.~\ref{eq:vband}).
This model reproduces the data well under many different experimental conditions, i.e., different lattice depths, atomic temperatures, post-cooling sequences, and Rabi-pulse durations. Notably, fits to the spectra using the Born-Oppenheimer and perturbed harmonic oscillator models yield different temperature estimates because the two approaches assume different relations between sideband frequency and atomic energy. This disagreement persists in the cold-atom regime and therefore cannot be attributed solely to the breakdown of the harmonic approximation at higher temperatures.

In the context of optical lattice clock accuracy, the choice of thermometry model affects both the inferred effective trap depths and the calculated lattice light shifts, with fractional-frequency differences of up to $8\times10^{-17}$. Lower atomic temperatures and the use of a consistent theoretical framework throughout the analysis reduce these differences. Yet, residual model dependence at the $10^{-18}$ level may still limit optical lattice clocks targeting uncertainties at this level with implications for high-accuracy frequency measurements such as precision clock comparisons~\cite{Godun2025,Pizzocaro2026}. These findings further motivate the refinement of sideband models beyond the approximations considered here, as well as the development of alternative thermometry techniques that are less sensitive to model-specific assumptions. Carrier-based thermometry has been demonstrated in several optical lattice experiments~\cite{McDonald2015,Han2018,Ranganath2026} and may provide a less model-dependent temperature estimate that can help clarify discrepancies between thermometry methods.

Although our study focused on Yb atoms, the methods and considerations presented here can be extended to other atomic species and to more general motional-state distributions, for example the preparation of selected longitudinal states~\cite{Nemitz2019, Bothwell2025a}. The Born-Oppenheimer model is particularly well suited to optical lattices, where a clear separation of motional energy scales naturally emerges, but it may also be applicable to other trapping platforms, such as optical tweezers~\cite{Kaufman2021} and atom chips~\cite{Keil2016, Bernon2013}.

\section*{Acknowledgement}
We thank F. Famà for carefully reading the manuscript and for insightful discussions.

This work was carried out in the project 22IEM01 TOCK. This project has received funding from the
European Partnership on Metrology, co-financed from the European Union’s Horizon Europe Research and Innovation Programme and by the Participating States.

This work was supported by the Cascading Grant Spoke 3 under the PNRR MUR project PE0000023-NQSTI, funded by the European Union – Next Generation EU.

\appendix
\section{Numerical calculations}\label{sec:numerical}
We evaluate Eq.~\ref{eq:vband} numerically using the open-source implementation available in Ref.~\cite{vbandGitHub}. The calculations are based on the potential energy curves $U_{n_z}$, which for the full potential can be calculated from Mathieu characteristic values~\cite{Beloy2020}. We based the evaluation of Mathieu special functions on the GNU Scientific Library (GSL)~\cite{Galassi2009}. The turning point $r_c=R_{n_z}(E)$ is obtained by numerically inverting $U_{n_z}(r)$. 

One possible strategy for calculating Eq.~\ref{eq:vband} is to approximate the energy integral by its dominant contribution at the resonant energy satisfying \(\delta=\delta\nu(E)\), similarly to the treatment in Ref.~\cite{Blatt2009}. This requires the inversion of Eq. \ref{eq:bo-freq}, which, for the full lattice potential, has to be done numerically.
Instead, we found it computationally faster to evaluate the energy integral directly on a finite grid for each $n_z$. The number of grid points can be chosen as a compromise between computational cost and numerical accuracy. In practice, the number of points is reduced with increasing $n_z$, because higher longitudinal states are less populated and have a smaller effect on the calculated line shape. The summation over $n_z$ is truncated by considering only trapped states or $U_{n_z}(0) < 0$.
%In practice, we reduce the number of points for higher $n_z$, as the contribution of higher level is reduced by the temperature. 

Eq.~\ref{eq:rabi} requires the integration of longitudinal wavefunctions $\mathcal{Z}_{n_z}$, which for the full potential can be calculated from Mathieu functions~\cite{Beloy2020}.
For computational speed we instead evaluate Eq.~\ref{eq:rabi} using the harmonic oscillator wavefunctions, even when the remaining quantities of Eq. \ref{eq:vband} are calculated using the full lattice potential.
In this approximation, defining the dimensionless local lattice depth $\mathcal{D}(r)=\frac{D}{E\ped{R}}e^{-\kappa^2r^2}$,
and the corresponding Lamb-Dicke parameter $\eta(r)=\frac{k\ped{c}}{\sqrt{2}k}\mathcal{D}(r)^{-1/4}$,
the Rabi frequency for a transition $n_z\rightarrow n_z+\Delta n_z$, with $\Delta n_z \geq 0$, is then~\cite{Wineland1979}
\begin{equation}
\frac{\Omega_{n_z,n_z+\Delta n_z}(r)}{\Omega_0}
=
e^{-\eta(r)^2/2}
\sqrt{\frac{n_z!}{(n_z+\Delta n_z)!}}\,
\eta(r)^{\Delta n_z}
L_{n_z}^{(\Delta n_z)}
\left[\eta(r)^2\right],
\end{equation}
where $L_n^{(\alpha)}$ is a generalized Laguerre polynomial, $k$ is the lattice wavevector, $\kappa = \sqrt{2}/w$, and $w$ is the waist of the lattice. In the turning-point calculation, this expression is evaluated at $r=r\ped{c}$.

\begin{figure}[h]
    \centering
    \includegraphics[width=0.5\linewidth]{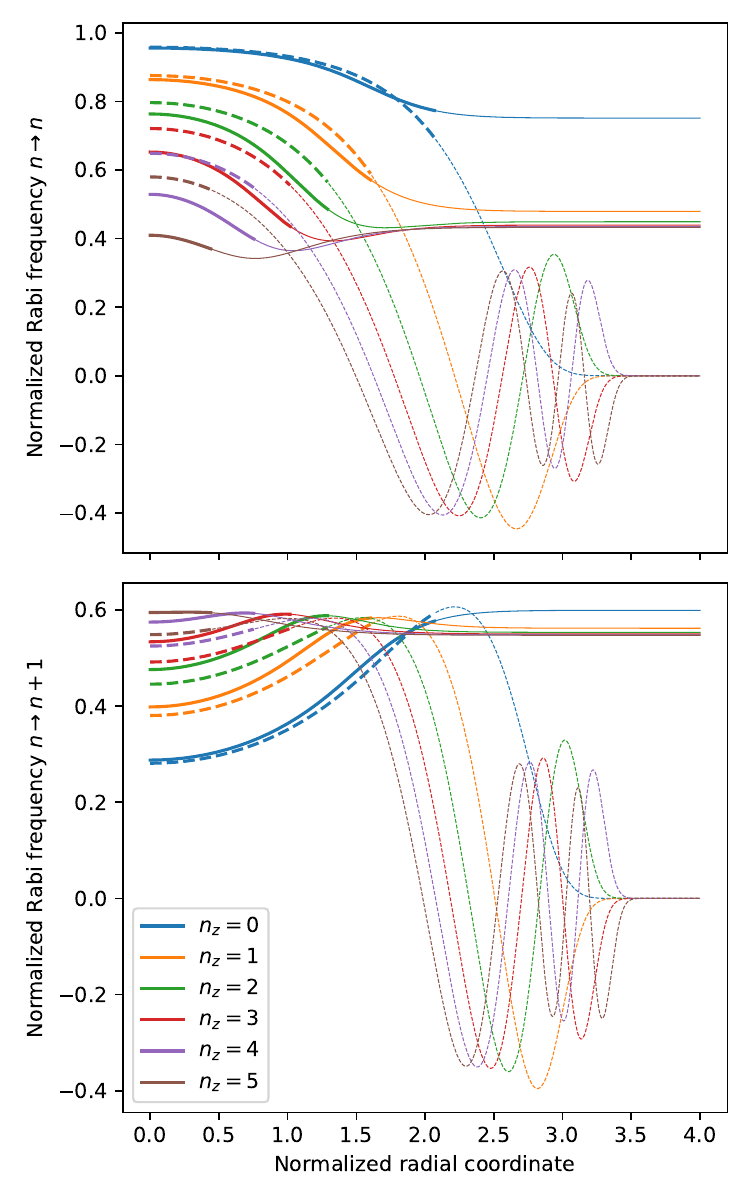}
    \caption{Comparison of the Rabi frequencies of Eq.~\ref{eq:rabi} calculated for the carrier $n_z \rightarrow n_z$ (upper panel) and first sideband $n_z \rightarrow n_z +1$ (lower panel) for a lattice trap depth of $D = 100\,E\ped{R}$ as a function of normalized radial coordinate $\sqrt{2} r\ped{c}/w$, where $w$ is the $1/e^2$ radius of the lattice. Color corresponds to different $n_z$ as in the legend. Full lines correspond to the exact Rabi frequency calculated using the solution to the axial Born-Oppenheimer equation calculated from Mathieu functions. Dashed lines correspond to approximated frequencies calculated for the perturbed harmonic oscillator. Very thin lines extend the calculations for regions that do not correspond to trapped atoms for the Born-Oppenheimer solutions, while dotted lines are the extended calculations for regions that do not correspond to trapped atoms for perturbed harmonic oscillator solutions.}
    \label{fig:rabi}
\end{figure}

As shown in Fig.~\ref{fig:rabi}, the two methods give similar Rabi frequencies for trapped atoms, both for carrier ($n_z \rightarrow n_z'=n_z$) and sideband ($n_z \rightarrow n_z'=n_z+1$) transitions.
In practice, since Eq.~\ref{eq:vband} includes many saturated, power-broadened transitions, the Rabi frequencies of the saturated transitions only determine the widths of the individual Lorentzian components. Therefore, the differences shown in Fig.~\ref{fig:rabi} have little effect on the overall sideband shape.
The harmonic oscillator Rabi frequencies are therefore sufficient to account for power broadening while substantially reducing the computational cost related to integrals of Mathieu functions. 
The free-space Rabi frequency $\Omega_0$ can be extracted from the carrier signal and kept fixed when fitting the sidebands.

\newpage

%\bibliography{./main.bib}

\end{document}